\RequirePackage[2020-02-02]{latexrelease}
\documentclass[PRL,preprintnumbers,floatfix,aps,notitlepage,nofootinbib,twocolumn,showpacs,amssymb]{revtex4-1}
\usepackage{amsmath,amsfonts,bm}
\usepackage{graphicx}
\usepackage{cancel}
\usepackage[colorlinks, linkcolor={red},citecolor={blue}]{hyperref}
\usepackage{xcolor}

\newcommand{\rs}{r_{\rm s}}

\begin{document}
\title{Black holes without Cauchy horizons and integrable singularities}
%
%
%
%
%
\author{J. Ovalle}
\email[]{jorge.ovalle@physics.slu.cz}
\affiliation{Research Centre for Theoretical Physics and Astrophysics,
	Institute of Physics, Silesian University in Opava, CZ-746 01 Opava, Czech Republic.}

\begin{abstract}
	In general relativity, nonsingular black holes contain (at least) a Cauchy horizon, a null hypersurface beyond which determinism breaks down. Even though the strong cosmic censorship conjecture establishes the impossibility of extending spacetime beyond this region, in this paper we investigate how far we can go, without invoking this conjecture, in the building of a physically reasonable black hole without a Cauchy hypersurface. Following this reasoning, we find a black hole lacking of Cauchy horizon, asymptotically flat and satisfying either the strong or dominant energy condition. The above is possible by demanding integrable singularity for the Ricci scalar, whose direct consequence is the appearance of finite tidal forces. {We show that the spacetime inside the event horizon represents a warped anti-de Sitter spacetime, which might be interpreted in terms of a finite superposition of configurations.} 
		\end{abstract} 
\maketitle
%
%
%
\section{Introduction}
One of the most notable achievements of general relativity (GR) is the successful prediction of black holes (BHs)~\cite{LIGOScientific:2016aoc,EventHorizonTelescope:2019dse}. It also predicts, under a set of physically reasonable conditions, the appearance of singularities as final stage of gravitational collapse, which, as the weak cosmic censorship conjecture (CCC) states, are hidden inside an event horizon, thus preventing that naked singularities can be seen by external observers~\cite{Penrose:1969pc,Hawking:1973uf}. However, the very existence of singularities, naked or dressed by an event horizon, should be a clear signal that GR has already reached its extreme domain, and therefore is falling apart. It is generally believed that such a problem would be overcome by quantum gravity, even though its complete formulation is still unknown.

Although we cannot understand the true nature of singularities in the framework of GR, it is possible to evade them by following a fairly simple strategy: generate regular BHs by filling the spacetime around the central singularity with some physically reasonable source of matter. This has produced a plethora of new regular BH solutions in recent years, mainly because the matter source used to evade the central singularity can be interpreted in terms of nonlinear electrodynamics. However, all these regular BH solutions contain a Cauchy horizon, a null hypersurface beyond which predictability breaks down, and also leads to mass inflation at the perturbative level~\cite{Poisson:1989zz,Poisson:1990eh}, a pathology which occurs even in loop quantum gravity inspired models~\cite{Brown:2011tv}. Regarding mass inflation, some important advances have been made recently~\cite{Bonanno:2020fgp,Carballo-Rubio:2022kad,Franzin:2022wai,Bonanno:2022jjp}, showing that singularity regularization does not always result in exponential instabilities. However, the existence of the Cauchy horizon remains potentially problematic~\cite{Carballo-Rubio:2018pmi}. A possible solution would be to generate regular BHs without inner horizon, and indeed some progress has been done in this direction by using classical~\cite{Boos:2021kqe} and quantum arguments~\cite{Casadio:2022ndh,Casadio:2023iqt}. Of course, we can always invoke the strong CCC~\cite{Penrose:1969pc}, which essentially ensures the impossibility of extending spacetime beyond the Cauchy horizon as a continuous Lorentzian manifold. But this is precisely contrary to what we want to develop, i.e., to avoid the very existence of Cauchy horizons.

Regarding the problematic existence of two horizons, there are well-known particular configurations in GR where the interior and exterior horizon are merged. These are extremal BHs, whose astrophysical relevance is null, at least for solutions known so far. Despite this fact, we wonder whether it is feasible that a regular single-horizon BH could be the final result from astrophysical collapse of compact objects. The aim of this paper is to investigate this point, exploring the possibility of building an astrophysically realistic example of GR being able to cope in high-curvature domains. In this respect, and even though there is not any rigorous formulation, it is generally believed that in the framework of classical GR it is impossible to have a regular BH (i) without a Cauchy horizon, (ii) asymptotically flat and (iii) satisfying either the strong or dominant energy condition. So far no counterexample is known, reinforcing the belief that such configurations cannot exist. In this paper we show that a class of such BHs does indeed exist by 
demanding integrable singularity for the Ricci scalar~\cite{Lukash:2013ts}, whose immediate physical consequence is the existence of finite tidal forces everywhere, even around the center of the BH configuration. To accomplish the above, we will build a ``Cauchy horizon free'' 
mass function for a metric describing the interior of a BH. As consequence of Einstein's equations, we find that this configuration contains an anti-de Sitter ($AdS$) spacetime filled with a regular matter function $\hat{m}(r)$, which is precisely the responsible for eliminating the Cauchy horizon. 

\section{Kerr-Schild spacetimes}
%
\label{sec2}
%

Let us start from the standard Einstein field equations
\begin{equation}
	\label{efe}
	R_{\mu\nu}-\frac{1}{2}\, R\,  g_{\mu\nu}
	=
	\kappa\,{T}_{\mu\nu}\ ,
\end{equation}
with $\kappa=8\,\pi\,G_{\rm N}$ and $c=1$.
It is well known that the line element for all spherically symmetric and static spacetimes can be written as~\cite{Visser:1995cc}
\begin{equation}
	ds^{2}
	=
	-e^{\Phi(r)}\left[1-\frac{2m(r)}{r}\right]\,dt^{2}+\frac{dr^2}{1-\frac{2m(r)}{r}}+r^2d\Omega^2
	\ ,
	\label{metric}
\end{equation}
where $\Phi(r)$ is a metric function and $m(r)$ stands for the Misner-Sharp mass function, 
which measures the amount of energy within a sphere of areal radius $r$. A particularly interesting case is that where the metric function $\Phi(r)=0$. Under this condition
the line element~\eqref{metric} belongs to the so-called spacetimes
of the Kerr-Schild class~\cite{kerrchild}, which has been extensively studied
(see \textit{e.g.} Ref.~\cite{Jacobson:2007tj}). In this case Einstein field equations~\eqref{efe} become 
\begin{eqnarray}
	\label{sources}
	\epsilon=\frac{2{m}'}{\kappa\,r^2}\ ,\,\,\,\,\,\,p_r=-\frac{2{m}'}{\kappa\,r^2}\ ,\,\,\,\,\,\,p_\theta=-
	\frac{{m}''}{\kappa\,r}\ .
\end{eqnarray}
where the energy-momentum tensor 
\begin{eqnarray}
	\label{emt}
	T_\mu^{\,\,\nu}=diag[-\epsilon,\,p_r,\,p_\theta,\,p_\theta]\ ,
\end{eqnarray}
contains an energy density $\epsilon$, radial pressure $p_r$ and transverse pressure $p_\theta$. Notice a characteristic feature of Einstein equations in~\eqref{sources}, which is a direct consequence of $\Phi(r)=0$, namely, the linearity in (derivatives of) the mass function ${m}(r)$. Hence, any solution $m(r)$ of the system~\eqref{sources} can be coupled with a second one $\hat{m}(r)$ to generate a new solution $\bar{m}(r)$ as
\begin{eqnarray}
	\label{gdks}
	m(r)
	&\rightarrow & \bar{m}(r) = m(r)+\hat{m}(r)
	\ . 
\end{eqnarray}
The expression in Eq.~\eqref{gdks} represents a trivial case of the so-called gravitational decoupling~\cite{Ovalle:2017fgl,Ovalle:2019qyi}. 
\par
 Since  the  Einstein  tensor  in  Eq.~\eqref{efe} satisfies the contracted Bianchi identities  $\nabla_\mu\,G^{\mu}_{\,\,\nu}=0$, the source $T_{\mu\nu}$ must be covariantly conserved, i.e., $\nabla_\mu\,{T}^{\mu\nu}=0$, yielding 
\begin{eqnarray}
	\label{con111}
	p_r'=
	-
	\left[\frac{\cancelto{0}{\Phi}'}{2}+\frac{m-r\,m'}{r(r-2\,m)}\right]\cancelto{0}{\left(\epsilon+p_r\right)}\,\,\,\,\,
	+\,\,
	\frac{2}{r}\left(p_\theta-p_r\right)\ .
\end{eqnarray}
If the metric~\eqref{metric} with $\Phi(r)=0$ describes the interior of a stellar object, we should expect the density
decays monotonically from a maximum at the origin $r=0$. Hence, from the expressions in Eq.~\eqref{sources}, exactly the opposite will occur for the radial pressure, i.e.,
\begin{equation}
	\epsilon'=-p'_r<0\rightarrow\,p'_r>0\ ,
\end{equation}
and therefore, according to Eq.~\eqref{con111}, the interior must be anisotropic, with  the transverse pressure always satisfying $p_\theta>p_r$.
This is characteristic for any self-gravitating body whose interior is described by a Kerr-Schild spacetime. In this respect, we want to highlight a critical feature regarding a Kerr-Schild fluid under hydrostatic equilibrium, which can be read directly from the expression~\eqref{con111}. Let us start by considering an element of volume $d\tau\,=4\pi\,r^2\,dr$ located at $r$ and in equilibrium. We see that, contrary to what happens in a conventional fluid, there is no ``gravitational force" $\sim\,\left(\epsilon+p_r\right)$ acting on the fluid element $d\,\tau$. Instead, it experiences a pull towards the center as a consequence of radial pressure gradients $p_r'>0$, which is canceled by a push towards the surface, a gravitational repulsion, precisely caused by the anisotropy in the pressures $\sim\,\left(p_\theta-p_r\right)$.

Another critical aspect is the continuity of the metric~\eqref{metric} with $\Phi(r)=0$. If this describes the spacetime of both an inner and outer region of a stellar object of radii $\rs$, then, in order to smoothly joint both regions at $r=\rs$, the mass function $m(r)$ must satisfy
\begin{equation}
	\label{c1}
	m(\rs)=\tilde{m}(\rs)\ ;\,\,\,\,\,\,\,m'(\rs)=\tilde{m}'(\rs)\ ,
\end{equation}
where $\tilde{m}$ stands for the exterior mass function and $F(\rs)\equiv\,F(r)\big\rvert_{r=\rs}$ for any function $F(r)$. Expressions in Eq.~\eqref{c1} are the necessary and sufficient conditions for smoothly joint both mass functions $\{m(r),\,\tilde{m}(r)\}$ at $r=\rs$. From the expressions in Eqs.~\eqref{sources} and~\eqref{c1} we conclude a significant feature regarding Kerr-Schild spacetimes: both density and radial pressure are continuous at the boundary $r=\rs$, that is,
\begin{equation}
	\label{c2a}
\epsilon(\rs^-)=\epsilon(\rs^+)\ ;\,\,\,\,\,\,p_r(\rs^-)=p_r(\rs^+)\ .
\end{equation}
However, the pressure $p_\theta$ is in general discontinuous.

\section{Regularity without\\ Cauchy horizon}

\label{sec3}
A simple strategy to investigate the existence of regular BHs with a single horizon is coupling different configurations, as shown in Eq.~\eqref{gdks}, until achieving the construction of a convenient mass function, i.e., that where any potential Cauchy horizon has been removed. However, despite the unlimited number of couplings we may carry out, when we demand a set of physically reasonable conditions, the chance of success in such a construction is extremely limited. Following this strategy, we provide what we consider to be the simplest possible nonsingular line element, representing a spherically symmetric ultracompact configuration of radius $\rs$, and lacking any potential Cauchy horizon, namely, the metric~\eqref{metric} with $\Phi(r)=0$ and mass function

\begin{equation}
	\label{m}
	m^\pm(r)=\frac{r}{2}\left[1\pm\left[1-\left(\frac{r}{\rs}\right)^n\right]^{k}\right]\ ;\;\;\;\;\;\ r\leq\rs
\end{equation}
where $\{k,\,n\}$ are constants to be discussed later. A simple analysis of the causal structure of this metric shows that it represents a BH for $m^+(r)$ and a horizonless ultracompact configuration for $m^-(r)$. In this work we will analyze the BH solution, leaving the horizonless case for future investigation. At this point we wish to stress that we do not make any attempt to postulate any action from which to derive the metric~\eqref{metric} with $\{\Phi=0,\,m^\pm\}$. Quite the contrary, our goal is to prove the existence of BH without Cauchy horizons, being simultaneously more tolerable to the singularity problem. 
In this respect, we remark the metric generated by the mass function~\eqref{m} is not globally defined but confined to $r\leq\,\rs$, 
where the total mass $M$ of the system, defined as
\begin{equation}
	\label{Mmim}
	M\equiv\,m(\rs)=\frac{\rs}{2}\ ,
\end{equation}
is independent of $\{k,\,n\}$, and therefore the event horizon is equal to the Schwarzschild's radius $\rs=2\,M$.
The source $\{\epsilon,\,p_r,\,p_\theta\}$ generating the BH solution, i.e., the metric~\eqref{metric} with $\{\Phi=0,\,m^+\}$, is found by using the mass function $m^+$ in Eq.~\eqref{sources}. A simple analysis shows that a monotonic decrease of the density $\epsilon(r)$ with increasing radius is only possible for i) $k=1;$ $n\in[\,0,1\,)$ and ii) $k>1;$ $n\in(\,1,2\,]$. In this regard, even though we should expect violation of energy conditions~\cite{Martin-Moruno:2017exc}, we found that the strong ($k=1$) and dominant ($k>6$) are satisfied. 
For the strong energy conditions, we have
\begin{eqnarray}
	&&
	\cancelto{0}{{\epsilon}+{p}_{r}}+\,{p}_{\theta}=\frac{n(n+1)}{2\,r^2}\left(\frac{r}{\rs}\right)^n
	\geq
	0\ ,
	\\
	\label{strong01}
	&&
	{\epsilon}+{p}_{\theta}=\frac{4+(n+1)(n-2)\left(\frac{r}{\rs}\right)^n}{2\,r^2}
	\geq
	0
	\ ,
\end{eqnarray}
where the expressions in Eqs.~\eqref{sources} and~\eqref{m} have been used.
We can proceed in a similar way (with arbitrary $k$) for the dominant energy conditions, $\epsilon\,\geq\,0\ ;\epsilon\,\geq\,|{p}_{i}|\quad
\left(i=r,\theta\right)$, although the expressions are more involved.

We see that the parameters $\{k,\,n\}$ produce a rich and physically attractive setting well worth investigating. However, keeping in mind the importance of vacuum energy as a source for quantum effects, and the well-known existence of a de Sitter core in regular BHs, in this work we will only be interested in the case  $n=2$, whose physical interpretation is quite straightforward, as we will see below. First of all, notice that the case $\{k=1,\,n=2\}$ yields
\begin{equation}
	\label{scalar}
	R=-4\,\Lambda+4/r^2\ ,
\end{equation}
where $R$ and $\Lambda=3/\rs^2$ are respectively the scalar curvature and cosmological constant. The expression in Eq.~\eqref{scalar} corresponds to an anti-de Sitter ($AdS$) spacetime filled with some matter producing a scalar singularity at the origin. Following the expressions in Eqs.~\eqref{gdks} and~\eqref{m}, we find that the $AdS$ behind our solution was deformed by coupling $m_{\rm AdS}\rightarrow\,m_{\rm AdS}+\hat{m}$, where $m_{\rm AdS}$ and $\hat{m}$ are respectively the $AdS$ mass function and $\hat{m}=r$. We see that the mass function $\hat{m}$ producing the singularity in Eq.~\eqref{scalar} is perfectly regular. We conclude that the singularity displayed in Eq.~\eqref{scalar}, necessary to remove the Cauchy horizon, is indeed integrable, where tidal forces remain finite~\cite{Lukash:2013ts} [notice that $\int\,R\,d\tau={\rm finite}$ yields $\int\,\xi_i\,d\tau={\rm finite}$ for each diagonal element $\xi_i$ in Eq.~\eqref{emt}, since $R\,\sim\,T$]. Now we can safely say that $k>1$ measures deviations from the deformed $AdS$ in Eq.~\eqref{scalar}. Indeed, by simple expansion of the metric function, i.e.
\begin{equation}
+g_{tt}=1-k\left(\frac{r}{\rs}\right)^2+\underbrace{\frac{k(k-1)}{2!}\left(\frac{r}{\rs}\right)^4{+\, .\, .\, .}}_{\rm ``deformation''}
\end{equation}
we can identify the effective cosmological constant as $\Lambda_k=3\,k/\rs^2$.
Finally, we want to highlight a critical point regarding the role of $k$ as generator of the deformed $AdS$. Notice that so far $k>1$ is an arbitrary constant, namely, $k\,\in\mathbb{R}$.
However, when we impose $k\,\in\mathbb{N}$, the warped $AdS$ has a simple
and fairly straightforward interpretation [see Eqs.~\eqref{gdks} and~\eqref{m}]: it is a
{\it finite} superposition of configurations. In particular, we can write the energy density in Eq.~\eqref{sources} as
\begin{eqnarray}
	\label{super}
	\kappa\epsilon_k(r)&=&\frac{1}{r^2}+k!\sum_{n=1}^{k}(-1)^{n}\frac{2n+1}{n!(k-n)!}\frac{r^{2(n-1)}}{\rs^{2n}}\nonumber\\
	&=&\frac{1}{r^2}+k!(-\Lambda+\epsilon^{(2)}_k-...+\epsilon^{(k)}_k)	\ ,
\end{eqnarray}
showing a finite superposition of $k$ configurations $\epsilon^{(n)}_k$, where $\epsilon_{\rm AdS}\equiv\epsilon_1$ is the (negative) vacuum energy, namely, the cosmological constant $\Lambda=3/\rs^2$. Since some configurations $\epsilon^{(n)}_k$ are negatives, we can interpret~\eqref{super} as a superposition of fluctuations in the $\epsilon_{\rm AdS}$ vacuum state, something which is quite suggestive as reminiscent of quantum  fluctuations.

Clearly the solution generated by the mass function~\eqref{m} is not asymptotically flat. In order to build a proper solution globally defined, we start by exploring whether is possible a smooth transition at $r=\rs$ between the interior solution, given by the metric~\eqref{metric} with $\{\Phi=0,\,m^+\}$, and an exterior with $\{\Phi=0,\,\tilde{m}(r)\}$, where $\tilde{m}(r)$ is the mass function for $r\geq\rs$. First of all, notice that the interior radial pressure in Eq.~\eqref{sources} at the surface $r=\rs^-$ is given by
\begin{equation}
	\label{prs}
	p_r(\rs^-)=-\frac{1}{\kappa\,\rs^2}\neq\,0\ .
\end{equation}
Hence, according to the matching condition~\eqref{c2a}, we conclude that the metric~\eqref{metric} with $\{\Phi=0,\,m^+\}$ cannot be smoothly coupled with the Schwarzschild's exterior solution (vacuum)
\begin{equation}
	ds^{2}
	=-
	\left(1-\frac{2{\cal M}}{r}\right)dt^{2}+\frac{dr^2}{1-\frac{2{\cal M}}{r}}+r^2d\Omega^2
	\ .
	\label{Schw}
\end{equation}
Therefore, we need a new exterior solution for the immediate surrounding region $r\sim\rs^+$ which quickly goes to Schwarzschild's exterior. This will be a transient region between the metrics~\eqref{metric} with $\{\Phi=0,\,m^+\}$ and~\eqref{Schw}. To accomplish this, we will make use of a new exterior solution recently developed through the so-called gravitational decoupling~\cite{Ovalle:2020kpd}, namely, a ``tensor-vacuum'' (an analogy to scalar vacuum) which satisfies the strong energy condition, and whose mass function reads
\begin{equation}
	\label{m2}
	\tilde{m}(r)
	=
	{\cal M}-\alpha\frac{r}{2}\,e^{-2r/(2\mathcal{M}-\ell)}
	\ ,
\end{equation}
where $\mathcal{M}$ is the asymptotic mass (the total mass measure by an asymptotic observer), $\alpha$ measures deviations from Schwarzschild's solution, and $\ell$ a ``hair" which for now satisfies $0<\ell\leq2\,{\cal M}$ to ensure asymptotic flatness, and whose physical meaning will be explained below. At this stage it is important to emphasize the difference between ${\cal M}$ and $M$ in Eq.~\eqref{Mmim}. The former is the total mass of the configuration, and in general is extended beyond $r=\rs$, while the latter is the fraction of this total mass confined within the region $r\leq\rs$.

Using the mass functions~\eqref{m} and~\eqref{m2} in the matching conditions~\eqref{c1}, we find $\{\mathcal{M},\,\alpha\}$ in terms of $\ell$ as
\begin{eqnarray}
	\label{match5}
	4{\cal M}_{\pm}&=&2\rs+\ell\pm\sqrt{\ell^2+4\rs\,\ell-4\rs^2}\ ,
\\
&&\nonumber
\\
	\label{match6}
	\alpha&=&\left(\frac{{2\cal M}}{\rs}-1\right)e^\frac{2\rs}{2\cal M-\ell}\ ,
\end{eqnarray}
where $\ell\geq\,2\rs(\sqrt{2}-1)$. By using the matching condition~\eqref{match6},
we find the exterior mass function~\eqref{m2} as
\begin{equation}
	\label{m3}
	\tilde{m}(r)={\cal M}+\frac{r}{2}\left(1-\frac{{2\cal M}}{\rs}\right)	e^{-2(r-\rs)/(2\mathcal{M}-\ell)}\ ,
\end{equation}
where the mass ${\cal M}$ in Eq.~\eqref{match5} has two branches $\{{\cal M}_{+},\,{\cal M}_{-}\}$ covering, respectively, ${\cal M}\geq\sqrt{2}\rs/2$ and ${\cal M}\leq\sqrt{2}\rs/2$. Plugging the mass function~\eqref{m3} into the line element~\eqref{metric} with $\Phi(r)=0$, we obtain the metric for the region $r\geq
\rs$ as
\begin{eqnarray}
	ds^2 =- &&\left[1-\frac{{2\cal M}}{r}-\left(1-\frac{{2\cal M}}{\rs}\right)	e^{-2(r-\rs)/(2\mathcal{M}-\ell)}\right]dt^2\nonumber\\
	+	&&\left[1-\frac{{2\cal M}}{r}-\left(1-\frac{{2\cal M}}{\rs}\right)	e^{-2(r-\rs)/(2\mathcal{M}-\ell)}\right]^{-1}dr^2\nonumber\\
	+	&&r^2d\Omega^2
	\ .
	\label{mimickerexterior}
\end{eqnarray} 
By simple inspection of the metric~\eqref{mimickerexterior}, we see that it represents a black hole with horizon
\begin{equation}
	\label{horizon}
	\rs=2\,M\,=\,({\cal M}+\sqrt{\ell{\cal M}-{\cal M}^2})\ ,
\end{equation}
and primary hair $\ell$ satisfying ${\cal M}\leq\ell\leq2\,{\cal M}$. In order to elucidate the physical meaning of this parameter, notice that from Eq.~\eqref{horizon} we find
%
%
%
%
%
%
\begin{eqnarray}
	\label{limit}
	\ell\rightarrow\,2{\cal M}\,\,\Rightarrow\,{\cal M}\rightarrow\,M\ .
\end{eqnarray}
Hence, we can say that $\ell$ controls the amount of mass ${\cal M}$ contained within the trapped surface $r=\rs$. We see that $\ell=2{\cal M}$ in Eq.~\eqref{limit} represents a Schwarzschild's BH, where the total mass ${\cal M}$ is confined within the region $r\leq\rs$. However, this case is explicitly excluded by the matching condition~\eqref{c2a}, and therefore ${\cal M}\leq\ell<2\,{\cal M}$. Nevertheless, the case $\ell\sim\,2{\cal M}$, i.e., ${\cal M}\sim\,M$ deserve more attention, as we will discuss later.
We conclude by specifying the mass function
\begin{equation}
	\label{mass}
	m(r)=\left\{
	\begin{array}{l}
		\frac{r}{2}\left[1+\left[1-\left(\frac{r}{r_{\rm s}}\right)^2\right]^{k}\right]\ ;\,\,\,\,\,\,\,\,0\leq\,r\leq\rs
		\\
		\\
		{\cal M}+\frac{r}{2}\left(1-\frac{{2\cal M}}{\rs}\right)e^\frac{-2(r-\rs)}{(2\mathcal{M}-\ell)}\ ;\,\,\,\,\,\,r\geq\rs\ ,
		\end{array}
	\right.
\end{equation}
which through~\eqref{metric} with $\Phi(r)=0$ yields a BH lacking of Cauchy horizon. We emphasize this solution contains three parameters $\{{\cal M},\,k,\,\ell\}$, where $k$ drives discrete deformations undergone by the interior $AdS$, and $\ell$ a primary hair driving the matter dressing the horizon. 
As expected, its surface gravity $\kappa$ and Hawking temperature $T_H$ are zero. We may be tempted to ensure that the solution represents a cold remnant. However, the configuration extends beyond the horizon, where the aforementioned variables have finite values for $r\sim\,\rs^+$. Since we have a hairy solution, we need to specify the Lagrangian ${\cal L}$ associate with the primary hair $\ell$ to find the entropy $S$~\cite{Wald:1993nt,Visser:1993nu}. In any case, notice that
\begin{equation}
	\label{S}
	\frac{A_{\rm H}}{4}=\pi\,{\cal M}\left(\ell+2\sqrt{\ell{\cal M}-{\cal M}^2}\right)\ .
\end{equation}


\par
\section{Apparent singularity\\and final remarks}
\label{sec4}
\par
In order to discuss an apparent nonintegrable singularity at the event horizon, let us analyze in detail what happens on the surface $r=\rs^+$. Since in general $m''$ is not continuous, there will be a discontinuity in the transverse pressure [see Eq.~\eqref{sources}], whose expression for the outer region is given by
\begin{equation}
	\label{pext}
	p^+_\theta(r)=\frac{2(r+\ell-2{\cal M})({\cal M}-M)}{\kappa\,M\,r\,(2\mathcal{M}-\ell)^2}\,e^\frac{-2(r-\rs)}{(2\mathcal{M}-\ell)}\ ,
\end{equation}	
We see that $p^+_\theta(\rs)\sim\, \frac{1}{2{\cal M}-\ell}$ for $\ell\rightarrow\,2{\cal M}\,\,({\cal M}\rightarrow\,M)$.
We conclude the greater the mass ${\cal M}$ packed in the region $r\leq\rs$, the greater the transverse stresses on the surface $r=\rs$, and indeed it diverges for ${\cal M}=M$ (which is excluded from our solution). Similarly, we find that curvature invariants on the horizon behave as
\begin{eqnarray}
	\label{curv}
	&&R\sim\frac{1}{{\cal M}}\frac{1}{2{\cal M}-\ell}\ ;\,\,\,\,R_{\mu\nu}R^{\mu\nu}\sim\frac{1}{{\cal M}^2}\frac{1}{(2{\cal M}-\ell)^2}\ ;\\
	\label{curv2}
	&&R_{\mu\nu\rho\sigma}R^{\mu\nu\rho\sigma}\sim\frac{1}{{\cal M}^2}\frac{1}{(2{\cal M}-\ell)^2}\ .
\end{eqnarray}
Although the case $\ell=2{\cal M}$ is excluded from~\eqref{mass}, it is worth investigating a possible solution without $(2{\cal M}-\ell)^{-1}$ in the curvature invariants. This is accomplished by demanding continuity of $m''$, which leads to $\ell={\cal M}$. In this case we end up with a simpler BH with only two parameters, i.e., $\{{\cal M},\,k\}$, where $\ell$ gives way to a secondary hair that we can read directly from Eq.~\eqref{mass}. However, only half of the total mass ${\cal M}$ will be enclosed by the horizon [see Eq~\eqref{horizon}], with the other half filling the region $r>\rs$, something whose phenomenological feasibility could mean quite a challenge. Therefore, in order to control the amount of mass contained within $r=\rs$, we need to keep alive the primary hair $\ell$. 



Coming back to our solution~\eqref{mass}, the expressions in Eqs.~\eqref{curv} and~\eqref{curv2} clearly indicate that the horizon $\rs=2\,M$ must be dressed with an amount of matter $d\,{\cal M}={\cal M}-M\neq\,0$ that cannot cross $r=\rs$, otherwise a naked singularity arises.  This process occurs thanks to the gravitational repulsion [see Eq.~\eqref{con111}] and seems to indicate a sort of exclusion mechanism to prevent singularities and other pathologies arising in classical GR. 
On the other hand, a phenomenologically viable BH should be that where almost all of ${\cal M}$ in the metric~\eqref{mimickerexterior} is contained in the region $r\leq\,\rs$, i.e., $d\,{\cal M}={\cal M}-M\sim\,0$, but this seems to bring us too close to the potential singularity displayed in Eqs.~\eqref{curv} and~\eqref{curv2} . This leads us to wonder whether it is possible to dress the black hole with a phenomenologically feasible amount of matter $d\,{\cal M}$ while being simultaneously far enough from the singularity.
A simple analysis shows a reasonable behavior of the solution for the case ${\cal M}/M\sim\,1.01$, i.e., where most of the total mass ${\cal M}$ ($\sim\,99\%$) is contained inside the trapped surface $r=\rs$. 
Such a small amount of matter $d\,{\cal M}\sim\,1\%$ is consistent with the fact that a perfect vacuum surrounding a BH is nothing but an idealized physical setting. 
Also notice that the more massive the object, the more tolerance it has for the potential divergence displayed in Eqs.~\eqref{curv} and~\eqref{curv2}. Hence, supermassive BHs only need a fairly small fraction of matter covering their horizon.
Finally, some aspects of the solution should be analyzed in depth, such as its stability, possible rotational extension and observational consequences, as well as time-dependent formation and evaporation, and most importantly, the action generating this solution. In this regard, an important guide should be the anisotropic effects causing the gravitational repulsion, necessary for noncollapsing matter, and critical for the existence of finite tidal forces. However, those are quite another matter beyond the target of this paper.

\subsection*{Acknowledgments}
This work is partially supported by ANID
FONDECYT Grant No. 1210041.
%

%
%
%
\bibliography{references.bib}
\bibliographystyle{apsrev4-1.bst}
%
%
\end{document}